\documentclass[11pt, a4paper]{article}
\usepackage{jheppub}
\usepackage{amsmath,amssymb,amsfonts}
\usepackage{longtable}
\usepackage{tikz}

\def\mylist{\begin{list}{}{\setlength{\leftmargin}{0.5in}
               \setlength{\listparindent}{-0.5in}
               \setlength{\itemindent}{\listparindent}}}

\newcommand{\bra}[1]{\langle#1\,|}
\newcommand{\ket}[1]{|#1\,\rangle}

\newcommand{\nD}[1]{\not D}

\title{Einstein Equations, Cosmological Constant and all that.}

\author{Denis Bashkirov}

\affiliation{Institut de Physique Th´eorique Philippe Meyer\\
Laboratoire de Physique Th´eorique, Ecole Normale Sup´erieure´\\
PSL Research University,\\
24 rue Lhomond, 75231 Paris Cedex 05, France $\&$\\
}

\emailAdd{denis.bashkirov@lpt.ens.fr}

\abstract{We suggest an interpretation of Einstein Equations of General Relativity at large scales in which the Cosmological constant is exactly zero in the limit of zero spacetime variations of fundamental constants. We argue that in a quasiclassical Universe such variation should be tiny which leads to a tiny value for the Dark Energy. Next, we suggest that the are two sources of the Dark Energy. The first is the variation in Newton's constant $G_N$. It is a form of Dark Energy in that it has negative pressure, but it differs from the Cosmological Constant by a negative contribution to the energy.  The second is the contribution of (causal) nonlocalities to the Dark Energy.

This comes together with a particular view of Quantum Mechanics and the wavefunction collapse, in particular. The collapse is neither dynamical nor subjective.}

\usetikzlibrary{trees}

\begin{document}

\maketitle

\flushbottom

\section*{Introduction}

We start with the following idea: suppose that the Universe is described by a (UV complete) Quantum Field Theory, and that it is in one of the simplest states one can have in a QFT -- states that are parameterized only by the metric/causal structure and the coupling constants/functions:
\begin{align}
\ket{Univ}=\ket{g_{\mu\nu}(x), c^a(x)}.
\end{align}

Then the expectation value of a stress tensor will be a causal nonlocal functional of these background functions
\begin{align}
\bra{Univ}\widehat T_{\mu\nu}(x)\ket{Univ} = T_{\mu\nu}[g_{\lambda\rho},c^a](x)\label{Tglob}
\end{align}
meaning that it only depends on their values in the past lightcone $J_-(x)$ of $x$.

In such a time-dependent background there will be particle production, so that the same expectation value will have another, effective, local as opposed to global (\ref{Tglob}), description in terms of particles and fields, so that
\begin{align}
T_{\mu\nu}[g,c](x)=T_{\mu\nu}(particles(x), fields(x)),\label{NLEE}
\end{align}
where the form of the representation on the right hand side is canonical. For example, its part corresponding to a ball of mass $m=1kg$ moving with four-velocity $u_\mu(x)$ is
\begin{align}
T_{\mu\nu}(ball(x))=mu_\mu(x)u_\nu(x)\approx (N_nm_n+N_pm_p)u_\mu(x)u_\nu(x)\nonumber\\
\approx 0.6\times10^{27}m_pu_\mu(x)u_\nu(x)\label{NLEEC},
\end{align}
and not anything else.

Because we identify $\ket{g,c}$ in the limit $\ket{g_{\mu\nu}(x)\to\eta, c^a=const}$ with the usual Minkowski vacuum in which $T_{\mu\nu}(prts, flds)=0$, the Cosmological constant is automatically set to zero.

A crucial point needs to be stressed right away. The relation (\ref{NLEE}) together with (\ref{NLEEC}) obviously constrains the possible geometries, while in the case of the usual particle production by varying background sources, for us the varying metric, the source/the metric is arbitrary. Our explanation of this is the central part of the conjecture.

It is as follows. The relation

\begin{align}
T_{\mu\nu}[g,c](x)=T_{\mu\nu}(particles(x)),\label{NLEE2}
\end{align}
does not constrain the metric in the context of small adiabatic quasi-homogeneous perturbations $\delta g_{\mu\nu}(x)$ on a quasi-homogeneous adiabatically changing background $g_{\mu\nu}(x)$ because the change of the left hand side when we pass from $g_{\mu\nu}(x)$ to $g_{\mu\nu}(x)+\delta g_{\mu\nu}(x)$ can be reproduced by small variations in the particles' state (it is a density matrix, but for notational simplicity we use ket-vectors) -- we change, for example, the coefficients in the superposition of local density eigenstates in the effective/particle description:
\begin{align}
 \ket{paricles(x)} = c_0(x)\ket{0(x)}+c_1(x)\ket{1(x)}+...+c_N(x)\ket{N(x)}+... .
\end{align}

So, infinitesimally, we can do this. The nontrivial question is whether this is possible for large variations of the metric on a generic, 'irregular' background.

{\it Our conjecture is that there are no (necessarily large) variations of the metric necessary to reproduce macroscopic deviations from the classical (approximate course-grained energy-momentum eigenstates) states of macroscopic bodies. In other words, if at an initial moment a ball of mass 1 kilogram is in a classical state $\ket{here, N\gg1}$, at a later moment it will remain in a (possibly different) classical state (with a small incertitude in position and particle number) no matter what we do to it\footnote{We could try to create a macroscopic superposition state by realizing the ball as the pointer of a measuring device coupled to a Schrodinger cat microscopic state a la Page and Geilker.}. Its macroscopic superpositions do not pass the matching condition for any metric.}

Since small arbitrary variations of the metric are still allowed, one can ask how deterministic Schrodinger evolution for small isolated systems is possible. Naively, there will be microscopically different metrics which would produce different unitary evolutions from a fixed initial state. 

Presumably, however, the corresponding variations in the metric are not homogeneous and slow enough for any noticeable indeterminism. Let us consider this situation in some more detail.

The right hand side of the condition \ref{NLEE2} does depend on the metric. For example, for a collection of free particles the Lagrangian is

\begin{align}
L = \sum_i\frac{1}{\sqrt{|g(x_i(\tau))|}}g_{\mu\nu}(x_i(\tau))\frac{dx^\mu(\tau)}{d\tau}\frac{dx^\nu(\tau)}{d\tau}.
\end{align}

Now, a microscopically small {\it isolated} system, due to the large value of the Planck constant (which is a function of the QFT parameters), will depend on it very weakly. Indeed, when we consider microscopic systems we never take the gravitational backreaction into account. So, the matching between the left and the right hand sides (of the nonlocal relation!) will be done iteratively: one starts with the right hand side and takes the metric in the particles' Hamiltonian/Lagrangian to be flat and computes the expectation of the stress tensor and a number of stress-tensor correlators (so that the matching is really at the quantum level). Then, one looks for the metric for the left hand side to reproduce the right hand side, finds it, plugs into the particles' Hamiltonian, computes small corrections to the RHS, then finds correction to the metric, etc. The validity of this iteration procedure will lead to the uniqueness of the solution to the matching equation.

One gets indeterminism and the freedom of choice for the metric when the iteration procedure breaks down. This happens during the measurement because of the involvement of a macroscopic measuring device, and in the adiabatic particle production. In the latter case, due to the violation of 'isolation' condition. In general, we expect the onset of this violation earlier (for smaller/lighter systems) than would be expected from the usual local Einstein equations.

So, we understand the condition of quasiclassicality of the Universe as the property that starting at some space and time scales and up, the effective reduced density matrix $\rho_{eff}(prts,flds)$ for a matter distribution in question computed from $\ket{Univ}$ in the usual way by taking the trace over the environment is strongly peaked around a (spacetime-averaged) $\widehat T_{\mu\nu}(x)$ eigenstate. In particular, there are not any macroscopic superpositions of position eigenstates.

In this situation the 'duality' relation (\ref{NLEE}) becomes a nonlocal version of General Relativity. In quasiclassical situations the nonlocal expression $T_{\mu\nu}[g,c](x)$ can be approximated by a local one. An heuristic argument is the following. The nonlocality gives rise to the causal propagator for $T_{\mu\nu}$ at separate times which is nothing but time ordered commutator of the stress tensor. In quasiclassical description one works with stress tensors averaged over macroscopic volumes, but these approximately commute since in the extreme volume limit we have $[P_\mu(t),P_\nu(t')]\approx 0$ -- in the quasiclassical situation the stress tensor is approximately conserved.  In the second part of the paper we investigate some microscopic obstacles to the existence of a local approximation. One obstacle for such an approximation to be natural (so that no huge coefficients are needed) is the conformal anomaly ${\cal A}[g,c]$. We find that a necessary condition on the QFT is a large vacuum expectation value $v$ for some operator with dimension close to one, which is identified with the Planck constant $M_P=v$.

Note that at the quasiclassical scale at which the geometry and the matter are related through the (nonlocal) Einstein Equations (\ref{NLEE}) the determinism does not arise in general due to the fact that in QFT the metric/causal structure is fundamentally not fixed by the theory -- it is a free parameter. Given an initial value of the metric and of the matter distribution $(g_i,m_i)$ at the quasiclassical scale, related by (\ref{NLEE}), one cannot always predict their future values $(g_f,m_f)$ even though they are also related by (\ref{NLEE}). In particular, this happens if in between the times $t_i$ and $t_f$ there is a 'contamination' from the microscale as during the measurement in Quantum Mechanics. Needless to say, this is a contentious idea.

In the first part of the paper we argue that for a Universe to exhibit quasiclassical behavior the spacetime variation of fundamental constants should be very small. There are no such restrictions on the variation of the metric $g_{\mu\nu}$ and background sources $A_\mu$ of conserved currents $\widehat J_\mu$ as well as vacuum expectation values $v(x)$. In the presence of a violation of the conservation of $\widehat J$ and an imperfect tuning of the vev $v$, an hierarchy between their variations (and also a constant value of $A_\mu$) and that of the metric appears.

In the second part we discuss the nonlocal Einstein Equations, their local approximation and the effect of time variation of the Planck mass. At the very end of the paper we speculate on the implications of our results for the interpretation of Quantum Mechanics.

\section{Quasiclassical variables}

We start with a general discussion of natural Quasiclassical variables in Quantum Field Theory irrespective of the global state and find state-independent restrictions on the QFT itself to exhibit quasiclassical behavior in any state.

Gell-Mann and Hartle  argued (\cite{GMH})\footnote{See especially \cite{H1} for a lucid explanation.} that conserved or approximately conserved quantities like the stress tensor and conserved currents corresponding to various symmetries naturally provide a set of quasiclassical variables -- the corresponding charges averaged over some space volumes $V$:
\begin{align}
\epsilon_V(t,{\bf y})\equiv\frac{1}{V}\int_{\bf y}d^3xT^{00}(t,{\bf x}),\nonumber\\
\pi^i_V(t,{\bf y})\equiv\frac{1}{V}\int_{\bf y}d^3xT^{0i}(t,{\bf x}),\nonumber\\
\nu_V(t,{\bf y})\equiv\frac{1}{V}\int_{\bf y}d^3xj^{0}(t,{\bf x}).
\end{align}

For completeness we briefly summarize their arguments.

A quasiclassical variable must correspond to a volume average of some quantum local operator $\widehat{\cal O}$. A necessary condition for quasiclassicality is that this variable exhibits a strong correlation in its values over time. Conserved charges have exactly this property. Thus one may expect that their densities integrated over large volume will have this tendency as well.

Furthermore, if one works in Heisenberg picture where the state does not change with time but local operators do, then the quasiclassical state at the initial and final times must be an approximate eigenstate of the two operators $\widehat{\cal O}(t)$ and $\widehat{\cal O}(t')$. For this to make sense the operator ${\cal O}(x)$ must approximately commute with itself at separate times -- so that they have common approximate eigenstates. For a random operator, this is just not true.

But because of approximate conservation of charges, $[\widehat P_\mu(t),\widehat P_\nu(t')]\approx 0$ and $[\widehat Q(t), \widehat Q(t')]\approx 0$ meaning that when the corresponding currents are averaged over sufficiently large but finite volumes $V$, they still approximately commute
\begin{align}
[\widehat T^V_{\mu\nu}(x),\widehat T^V_{\rho\lambda}(y)]\approx 0,\nonumber\\
[\widehat J^V_\mu(x),\widehat J^V_\nu(y)]\approx 0,\nonumber\\
[\widehat J^V\mu(x),\widehat T_{\rho\lambda}(y)]\approx 0.
\end{align}
whenever $x=(x^0,{\bf x})$ and $y=(y^0,{\bf y})$ belong to different cells in this 'cell' decomposition of the spacetime. In the rest of the section we suppress the superscript ${}^V$ in order not to clutter the formulae, but all local operators are assumed to be averaged over these sufficiently large volumes.

That they approximately commute, means that we can talk about their approximate common eigenstates $\ket{T_{\mu\nu}(x),J_\rho(x)}$.

For the rest of this section we confine our attention to the stress tensor and assume that there are no other (quasi)conserved currents in the theory.

We will not assume that averaging over spatial volumes is enough, and will permit some smearing in the temporal direction as well.

The requirement of the quasiclassical behavior of $\widehat T_{\mu\nu}(x)$ in a state $\ket{\Psi}$ is twofold.
First of all, quantum fluctuations around some mean value $T_{\mu\nu}(x)$ must be negligible in some region of spacetime ${\cal M}$. It will prove useful later to reformulate this constraint in the following way: the state $\ket{\Psi}$ is an approximate eigenstate of the quantum operator $\widehat T_{\mu\nu}(x)$ in a spacetime region $x\in{\cal R}_{\cal M}$ with the approximate eigenvalue $T_{\mu\nu}(x)$:
\begin{align}
\widehat T_{\mu\nu}(x)\ket{\Psi}\approx T_{\mu\nu}(x)\ket{\Psi},\qquad x\in{\cal R}_{\cal M}.\label{eigr1}
\end{align}

That this is not a sufficient condition for the validity of a quasiclassical approximation is obvious from the fact that the stress tensor $\widehat T_{\mu\nu}(x)$ does not satisfy any dynamical equations. Correspondingly, although fluctuations around the mean value $T_{\mu\nu}(x)$ are negligible, the value itself is not determined from any initial data. The only equation that the stress tensor obeys is the conservation equation $\partial_\mu\widehat T_{\mu\nu}(x)=0$ which is a constraint condition on its components. This is a set of four equations on ten components of the stress tensor.

For the quasiclassical approximation to hold the quasiclassical variables must obey a set of deterministic dynamical equations. Thus there must be additional constraints on the state $\ket{\Psi}$ which effectively reduce the ten components of the stress tensor to four. Furthermore, the resulting equations must be deterministic in the sense that later evolution of the quasiclassical variable is uniquely determined by the initial data in simple enough situations.

The second condition on the state that one needs to impose to recover classical physics is, for example, the condition of applicability of local thermodynamics. This is the requirement that for a (necessarily small) set of local operators ${\cal O}(x)$ the expectation values $\bra{\Psi}{\cal O}(x)\ket{\Psi}$ and local correlators (with not too many insertions) are well approximated by the local density matrix
\begin{align}
\bra{\Psi}{\cal O}_1(x_1)...{\cal O}_k(x_k)\ket{\Psi}\approx tr[{\cal O}_1(x_1)...{\cal O}_k(x_k)\rho_{leq}[u]],
\end{align}
where\footnote{It actually should be the causal version of the given expression.}
\begin{align}
\rho_{leq}[u]=Z^{-1}\exp{[-\int d^3xu^\mu(x)\widehat T_{\mu 0}(x)]}.
\end{align}
 To make contact with the standard thermodynamic expression for $\rho_{leq}$ one needs to separate the norm of the four-vector from its direction $u^\mu$ by introducing the inverse 'temperature' $\beta(x)$ and normalization $u^\mu(x)u_\mu(x)=-1$ so that the density matrix takes the form
\begin{align}
\rho_{leq}[u]=Z^{-1}\exp{[-\int d^3x\beta(x)(\epsilon(x)-{u_i}(x){\pi^i}(x))]}.
\end{align}

In particular

\begin{align}
T_{\mu\nu}(x)\approx Z^{-1}tr[\widehat T_{\mu\nu}(x)\exp{[-\int d^3zu^\lambda(z)\widehat T_{\lambda 0}(z)]}]
\end{align}

The mean value on the left hand side becomes a functional of $u^\mu(z)$ in addition to being a function of $x$:

\begin{align}
T_{\mu\nu}(t',{\bf x})\approx T_{\mu\nu}[u^\lambda(t,{\bf y})](t',{\bf x}).\label{eigr2}
\end{align}

One assumes that the same effective density matrix $\rho_{leq}[u(t,{\bf x})]$ can be represented at a different time $t'$ by the same expression with a different function of ${\bf x}$
\begin{align}
\rho_{leq}[u(t,{\bf x})]=\rho_{leq}[u(t',{\bf x})].
\end{align}

Thus the expectation value of $T_{\mu\nu}(t',{\bf x})$ has a local form in terms of $u(t',{\bf x})$:
\begin{align}
T_{\mu\nu}(t',{\bf x})\approx T_{\mu\nu}[u^\lambda(t',{\bf y})](t',{\bf x})= mu_\mu(t',{\bf x})u_\nu(t',{\bf x})+p(u(t',{\bf x}))\eta_{\mu\nu}+... .
\end{align}
The conservation of the stress tensor becomes a hydrodynamical equation in this ansatz.

\section{Quasiclassicality in the presence of spacetime-dependent \\
background sources.}

Having reviewed the emergence of quasiclassical physics in QFT in the usual setup with flat spacetime background we move on to consider curved backgrounds which include a curved spacetime metric $g_{\mu\nu}(x)$ as well as spacetime dependent background sources $c(x)$ which are obtained by promoting coupling constants to spacetime functions.

While some of coupling constants may be forbidden from being functions on spacetime (if it leads, f.e., to violations of causality) rather than numbers, there is no general argument against such a promotion. Furthermore, from the point of view of QFT as quantum theories defined on the space of causal structures rather than on the space of Lorentzian metrics \cite{B}, the existence of a conformal frame in which all background sources are constants seems unnatural and requiring an explanation.

In this section we aim at providing a possible explanation of this fact. We argue that spacetime variations of fundamental constants lead to breakdown of quasiclassical approximation much faster than variations in the metric. Their observed constancy (compared to the metric) is therefore related to the fact that the Universe is extremely well described by classical physics. Thus the requirement of quasiclassicality introduces an hierarchy in the variations of the metric and other background sources.

Let us start with the modification of the conservation equation of the stress tensor for the case of switched-on background sources. In this section we consider only two kinds of background sources: the metric $g_{\mu\nu}(x)$ itself and a set of scalar coupling constants $\lambda^a(x)$ promoted to functions. The generalization of the stress tensor conservation equation is
\begin{eqnarray}
\nabla^\mu\widehat T_{\mu\nu}(x)+\partial^\nu c^a\widehat{\cal O}_a(x)=0.
\label{diffinv}
\end{eqnarray}
Here ${\cal O}_a(x)$ is the local operator associated to the coupling constant $c^a$ -- in perturbation theory one inserts $exp{[i\int d^4x\sqrt{|g|}c^a\widehat{\cal O}_a(x)]}$ in all correlation functions.

This equation expresses the diffeomorphism-invariance of the theory and is a constraint on operators, and not just their expectation values. For the purpose of brevity, to illustrate the connection to the diffeomorphism invariance we use the Schwinger generating functional for connected correlators (in flat space and constant fundamental 'constants') $W[g_{\mu\nu}(x), c(x);\mu]$. Under infinitesimal diffeomorphisms $\delta x^\mu=\xi^\mu$, $\delta g_{\mu\nu}=\nabla_{(\mu}\xi_{\nu)}$ and $\delta c^a=\partial^\mu c^a\xi_\mu$ so that
\begin{eqnarray}
0=\delta W=\int d^4x\sqrt{|g|}[\frac{\delta W}{\sqrt{|g|}\delta g_{\mu\nu}(x)}\delta g_{\mu\nu}(x)+\frac{\delta W}{\sqrt{|g|}\delta c(x)}\delta c(x)]\nonumber\\
=\int d^4x \xi^{\mu}[\nabla^\nu \langle T_{\mu\nu}(x)\rangle+\partial_\mu c^a\langle{\cal O}_a(x)\rangle].
\end{eqnarray}

Due to arbitrariness of $\xi^\mu$ this is equivalent to the identity
\begin{eqnarray}
\nabla^\nu \langle \widehat T_{\mu\nu}(x)\rangle+\partial_\mu c^a\langle\widehat {\cal O}_a(x)\rangle=0.
\end{eqnarray}
Here the bracket does not correspond to an expectation value in any state, but to an inner product of the in- and out- vacua $\ket{0_{in}}$, $\ket{0_{out}}$ in the presence of sources with nontrivial spacetime-dependence with the operators sandwiched between them \cite{BD,MW}:
\begin{eqnarray}
\nabla^\nu \bra{0_{out}} \widehat T_{\mu\nu}(x)\ket{0_{in}}+\partial_\mu c^a\bra{0_{out}}\widehat {\cal O}_a(x)\ket{0_{in}}=0.
\end{eqnarray}

Compared to the case of flat spacetime and constant fundamental sources, the diffeomorphism-invariance equation \ref{diffinv} contains additional terms of two kinds:
\begin{align}
\partial^\mu\widehat T_{\mu\nu}(x)=-\Gamma^\nu{}_{\mu\rho} \widehat T^{\mu\rho}(x)-\Gamma^{\mu}{}_{\mu\rho} \widehat T^{\nu\rho}(x)-\partial^\nu c^a\widehat{\cal O}_a(x).
\end{align}
The first two terms on the right hand side are proportional to the stress tensor $\widehat T^{\mu\nu}$ and the last one -- to the operators $\widehat{\cal O}_a(x)$. In a generic Lorentzian geometry close to the flat spacetime both kinds of terms are obstacles to the conservation of the stress tensor. In this case the stress tensor is only approximately conserved.

Naively, one would then expect spacetime variations of the metric and the sources $c^a(x)$ to be on equal footing with regards to the deterioration of the validity of the classical approximation. However, there is a very important difference between the two. Switching on a variation of the metric introduces the stress tensor into the right hand side of the equation, while switching on variation of the sources $c^a(x)$ introduces the operators $\widehat {\cal O}_a(x)$. The latter, unlike the former, do not satisfy any conservation constraints and thus their averaged values will not show strong correlations in time. Thus their presence in the conservation equation 'contaminates' the stress tensor itself. In other words, if initially at a moment $t_0$ the state $\ket{\Psi}$ is an approximate eigenstate $\widehat{\cal O}_a(t_0,{\bf x})$, it will be far from that for $\widehat{\cal O}_a(t>t_0,{\bf x})$, and hence the same becomes true for the stress tensor if there is a spacetime dependence in the corresponding coupling function.

In fact, the last statement requires some further scrutiny, as there is a possibility that the linear combination $\partial^\mu c^a(c)\widehat{\cal O}_a(x)$ can be expressed through the trace of the stress tensor $\widehat T_\mu^\mu$ and the identity operator for a special choice of nonconstant $c^a(x)$.

Indeed, the trace of the stress tensor in QFT is a linear combination of local operators explicitly breaking the conformal symmetry of the UV fixed point and the identity operator with the coefficient proportional to the conformal anomaly:
\begin{align}
\widehat T_\mu^\mu(x)=c^A(x)\widehat{\cal O}_A(x)+{\cal A}[g_{\mu\nu},c_a]\widehat 1.\label{trace}
\end{align}

Were the set of indices $\{A\}$ to coincide with the set $\{a\}$, the choice of variations of the sources satisfying the relation $\partial^\mu c^a(x)=\lambda(x)c^a(x)$ $\forall a$ for any function $\lambda(x)$ would transform the equation \ref{diffinv} into
\begin{align}
\nabla^\nu\widehat T_{\mu\nu}(x)+\lambda(x)(\widehat T_\mu^\mu(x)-{\cal A}[g,c]\widehat 1)=0.
\end{align}

In this case the variations of all fundamental 'constants' would be dictated by one arbitrary function $\lambda(x)$, and there would be no obstacle for the stress tensor to be a good quasiclassical variable.

Thus one needs to examine the conditions under which such a coincidence of sets $\{A\}$ and $\{a\}$ occurs. To do this let us turn again to the Schwinger functional $W[g_{\mu\nu},c^a;\mu]$. It is invariant under Weyl transformations up to the conformal anomaly when the sources transform appropriately together with the metric:
\begin{align}
\int d^4x\{\frac{\delta W}{\delta g_{\mu\nu}(x)}\delta_\sigma g_{\mu\nu}(x) +\frac{\delta W}{\delta c^a(x)}\delta_\sigma c^a(x)\}={\cal A}[g_{\mu\nu},c^a;\sigma;\mu].\label{wtrace}
\end{align}

As before, the variations of $W$ with respect to the sources give the matrix elements of the corresponding local operators between the in- and out- vacuum states $\ket{0_{in}}$ and $\ket{0_{out}}$.
\begin{align}
\int d^4x\{\bra{0_{in}}\widehat T_{\mu}^\mu(x)\ket{0_{out}} +\bra{0_{in}}\widehat{\cal O}_a(x)\ket{0_{out}}\sqrt{|g|}\delta_\sigma c^a(x)\}={\cal A}[g_{\mu\nu},c^a;\sigma;\mu].\label{wtrace2}
\end{align}

The variations $\delta_\sigma c^a$ of the sources are expected to have the most general form consistent with symmetries and with dimensional analysis.

Let us assume for a moment that among the coupling functions $c^a(x)$ there is at least one $m^2(x)$ with mass dimension two. Then its transformation law under the Weyl symmetry \cite{Ratal} will contain a derivative term
\begin{align}
\delta_\sigma m^2(x)\supset f(c^a)\Box_g\sigma(x),
\end{align}
which after substitution into equation \ref{wtrace} and integration by parts gives rise to the presence of the operator $\Box_g\widehat{\cal O}_{m^2}(x)$ in the expression for the trace of the stress tensor:
\begin{align}
\widehat T_{\mu}^\mu(x)\supset \Box_g\widehat{\cal O}_{m^2}(x).
\end{align}
At the same time the diffeomorphism invariance operator equation \ref{diffinv} contains operator $\widehat{\cal O}_{m^2}(x)$ but not its descendant $\Box_g\widehat{\cal O}_{m^2}(x)$. Accordingly, in this case the 'bad' operator entering the diffeomorphism equation is not expressed through the trace of the stress tensor and the identity operator and the corresponding sources $m^2(x)$ has to be very close to a constant in quasiclassical regime.

We will see later that the operator $\widehat{\cal O}_{m^2}(x)$ enters the story naturally as one of the relevant deformations of the UV fixed points, so we will not try to elucidate further the connection between the sets of operators entering the two equations \ref{diffinv}, \ref{trace}. Nevertheless, it is worth mentioning that the analysis presented in \cite{Ratal} appears incomplete as the mass parameter $\mu$ played no role in the transformation properties of the sources, which does not seem natural given that its appearance in QFT is exactly to make dimensional analysis work in a more general way than in classical physics. In other words, we expect $\mu$ to enter the transformations of the sources thus facilitating appearance of derivatives of $\sigma$ even in the absence of sources with integral conformal dimensions near the UV CFT.

Thus appearance of operators $\widehat{\cal O}_a(x)$ on the right hand side of the diffeomorphism invariance equation \ref{diffinv} is an obstacle for the stress tensor to be a quasiclassical variable and for the existence of the quasiclassical approximation itself. To avoid this one has to have a negligible spacetime variation of $c^a(x)$. Because there is no such problem with variations of the metric, one naturally gets a large hierarchy $L(\delta g_{\mu\nu})\ll L(\delta c^a)$ between the scales of variations of the metric $L(\delta{g_{\mu\nu}})$ and of other sources $L(\delta{c^a})$ in a quasiclassical regime.

\section{'Vacuum' state.}

In this section we define what we mean by the 'vacuum' state in arbitrary backgrounds (actually, in arbitrary backgrounds in some neighborhood of the flat spacetimes with constant sources). We will not be overly concerned with the mathematical question of existence of such vacuum states. Physics evidence of their existence will be discussed. We do not investigate the neighborhood of Minkowski space in which this family is defined but will make some comments on it.

We start with a formal definition and later add some concreteness.

There are two main aspects to our definition of 'vacuum' states: 1) they are defined not in 'isolation' but in relations to each other, so as a family (akin to categorical approach in mathematics), 2) there is a correspondence limit: the 'vacuum' states must agree with the standard definition on members for which there is one.

In a sense, it will be the most symmetric state in a given background.

Now let us explain what we mean by the vacuum on an arbitrary curved spacetime background $g_{\mu\nu}(x)$ with background sources/coupling functions $c^a(x)$.

1) A vacuum $\ket{g,c}$ is a physical state, so according to the standard definition, it is a positive linear functional on the $C^*$-algebra of (local) observables which to any product of local operators $\widehat{\cal O}_i(x_i)$ sets into correspondence its expectation value:

\begin{align}
\ket{g,c}:\qquad \prod_{i=1}^k\widehat{\cal O}_i(x_i)\to\bra{g,c}\prod_{i=1}^k\widehat{\cal O}_i(x_i)\ket{g,c}.
\end{align}

2) For any collection of local operators $\widehat{\cal O}_i(x_1)$ the expectation value viewed as a functional of the sources
\begin{align}
F_{\cal O_i}[g,c](x_i)\equiv\bra{g,c}\prod_{i=1}^k\widehat{\cal O}(x)\ket{g,c}
\end{align}
is only dependent on the value of sources in the causal past $J^-(\cup_{i=1}^kx_i)=\cup_{i=1}^kJ^{-}(x_i)$ of the insertion points $x_i$.

3) The correspondence limit: for the special case of Minkowski spacetime $g_{\mu\nu}(x)=\eta_{\mu\nu}$ and constant background sources $c^a(x)=c^a$ the vacuum functional gives the standard vacuum correlation functions in the theory with coupling constants $c^a$:
\begin{align}
\bra{\eta,c}\prod_{i=1}^k\widehat{\cal O}_i(x_i)\ket{\eta,c}=\bra{0}\prod_i\widehat{\cal O}_i(x_i)\ket{0}
\end{align}

As follows from this definition, any element of the Hilbert space obtained from a vacuum $\ket{g,c}$ by the action of local (smeared) operators will not itself be another vacuum state.

{\it Remark:} There is an interesting question of the common domain $D(g,c)$ of all functionals $F_{\cal O_i}[g,c](x_i)$. It is natural to expect that it at least includes all smooth sources $(g,c)$ whose deviations from the Minkowski metric with constant sources have compact support on the spacetime manifold or that are asymptotically flat in the past. In this paper we do not investigate this issue, and our statements will be valid for all members of $D(g,c)$ whatever it is.

Finally, we should stress that we work in the Heisenberg picture in which states do not change with time.

To make contact with the known definition of vacuum states on curved spacetime backgrounds and/or switched on spacetime-dependent sources, let us consider the situation when the spacetime metric is asymptotically flat in the past and the future, and the other sources approach constant values asymptotically.\footnote{For a pedagogical exposition see books \cite{BD,MW}}

In this situation there are two notions of the vacuum: the in-vacuum $\ket{0_{in}}$ is the state which looks like the standard Minkowski vacuum $\ket{\eta,c^a_{in}}$ with coupling constants  $c^a_{in}=\lim_{t\to-\infty}c^a(x)$, that is
\begin{align}
\bra{0_{in}}\prod_i{\cal O}(t_i,{\bf x}_i)\ket{0_{in}}\xrightarrow[t_i\to-\infty]{}\bra{\eta,c^a_{in}}\prod_i{\cal O}(t_i,{\bf x}_i)\ket{\eta,c^a_{in}},
\end{align}

and the out-vacuum $\ket{0_{out}}$ defined analogously:

\begin{align}
\bra{0_{out}}\prod_i{\cal O}(t_i,{\bf x}_i)\ket{0_{out}}\xrightarrow[t_i\to+\infty]{}\bra{\eta,c^a_{out}}\prod_i{\cal O}(t_i,{\bf x}_i)\ket{\eta,c^a_{out}}.
\end{align}

They are different states if $c^a_{in}\ne c^a_{out}$ or if the metrics varies in between. The in-vacuum $\ket{0_{in}}$ looks like the vacuum in the far past but contains particles in the future, while the out-vacuum $\ket{0_{out}}$ looks like the vacuum in the far future but contains particles in the past. Obviously, the latter state violates the second law of thermodynamics, so the preferred state of the two is the in-vacuum $\ket{0_{in}}$.

With our definition this choice is automatic as a consequence of the properties (2) and (3).

\section{Einstein Equations as a necessary condition of \\
quasiclassicality of the 'vacuum'.}

In cosmology as the initial state one often takes the vacuum state in some symmetric spacetime background, f.e, the Bunch-Davies vacuum $\ket{0_{BD}}$ at the neck of the de Sitter spacetime \cite{BD,MW}.

In this section we do not want to consider any symmetric spaces. Rather, given the definition of the vacuum in a general background from the previous section, we ask the question: for what background metric and other sources the vacuum state $\ket{g,c}$ exhibits quasiclassical behavior. That is, for what functions $(g_{\mu\nu}(x),c^a(x))$ the stress tensor $\widehat T_{\mu\nu}(x)$ is a good quasiclassical variable in the vacuum state $\ket{g,c}$.

One should keep in mind that the vacuum state $\ket{g,c}$ will look nothing like an empty space at finite times for generic sources due to the particle production, as was mentioned in the previous section. Generically, the vacuum state will contain a complicated bath of particles. The question is under what conditions on the background sources $(g_{\mu\nu}(x),c^a(x))$ the evolution of this bath will be described by classical equations with probabilities strongly peaked around solutions of these classical equations.

A part of the question was already answered in section 2. A necessary condition is a large hierarchy between spacetime variations of the metric and the other sources: $L(g_{\mu\nu})\ll L(c^a)$. This is true under the assumption discussed in the aforementioned section.

The more difficult part of the question to answer is: what are the requirements on the metric? Below we answer this question and find out that a necessary condition on the metric is its satisfying the Einstein Equations of General Relativity.

We already know that a necessary condition on the state is that it is an approximate eigenstate of the stress tensor operator (averaged over some volumes):
\begin{align}
\widehat T_{\mu\nu}(x)\ket{g,c}\approx T_{\mu\nu}(x)\ket{g,c},\qquad x\in{\cal R}_{\cal M}.
\end{align}
The deterministic classical equations appear only when further the expression $T_{\mu\nu}[u^\lambda(z)](x)$  is well approximated by a local expression in terms of particles and fields. For example, it could be a hydrodynamical description:
\begin{align}
T_{\mu\nu}(x)\approx T_{\mu\nu}[u^\lambda(z)](x)\approx mu_\mu(x)u_\nu(x)+p(u(x))\eta_{\mu\nu}+...=T^{hydro}_{\mu\nu}(x) .
\end{align}

On the other hand, from the definition of the vacuum, the expectation value of any local operator only depends on the background sources. So
\begin{align}
\bra{g,c}\widehat T_{\mu\nu}\ket{g,c}=F_T[g_{\mu\nu}(z),c^a](x).\label{Tmic}
\end{align}
In QFT there is an ambiguity in the choice of the stress tensor with any two choices differing by the variational derivative with respect to the metric of a local functional $Loc[g,c]$ (multiplied by the unit operator). This defines an equivalence class $[\widehat T_{\mu\nu}(x)]$ of all stress tensors: for any two representatives $\widehat T_{\mu\nu}(x)$ and $\widehat T'_{\mu\nu}(x)$ there exists a local functional $Loc[g,c]$ such that
\begin{align}
\widehat T_{\mu\nu}(x)-\widehat T'_{\mu\nu(x)}=\frac{\delta Loc[g,c]}{\sqrt{|g|}\delta g^{\mu\nu}}(x)\widehat 1.
\end{align}

Let us denote $U(g,c)\subset D(g,c)$ the set of all sources $(g,c)$ for which there exists a $\widehat T_{\mu\nu}(x)$ from the equivalence class $[\widehat T_{\mu\nu}(x)]$ that satisfies
\begin{align}
\widehat T_{\mu\nu}(x)\ket{g,c}\approx T_{\mu\nu}(prts(x),flds(x))\ket{g,c}
\end{align}
In particular, the expression on the right hand side may be a hydrodynamical expression for some vector field $u^\mu(x)$. Let us consider the subset in $U(g,c)$ which is the connected component $U_M(g,c)$ of the Minkowski case $(\eta, c=const)$. For example in the case of hydrodynamics with lumps of matter, starting with an element of $U_M(g,c)$ one can get continuously to the Minkowski case by keeping the spacetime support of lumps fixed by rescaling their 'height' to zero. In this connected component the expectation values of the microscopic $\widehat T_{\mu\nu}$s are given by a single analytic expression obtained as
\begin{align}
\bra{g,c}\widehat T(x)\ket{g,c}\approx\frac{\delta W_{loc}[g,c]}{\sqrt{|g|}\delta g(x)}.
\end{align}
The Minkowski case corresponds obviously to $u^\mu(x)\equiv 0$ (thus $T^{hydro}_{\mu\nu}\equiv0$), and hence the microscopic $<T>(\eta,c=const)\equiv 0$ which is the correspondence limit condition.

Thus there is a single stress tensor whose expectation value is given fundamentally by the formula \ref{Tmic}. Furthermore, as we argued in the introduction the quasiclassical regime is the one in which quantum nonlocalities (nonzero commutators of the averaged stress tensor at separate points) are neglected. So we are in a situation when the nonlocal functional \ref{Tmic} is approximated by the local one. In other words, it is approximated by the variational derivative of a local functional $W_{loc}[g,c]$:
\begin{align}
\bra{g,c}\widehat T_{\mu\nu}\ket{g,c}=F_T[g_{\mu\nu}(z),c^a](x)\approx\frac{\delta W_{loc}[g,c]}{\sqrt{|g|}\delta g^{\mu\nu}(x)}.
\end{align}

Equating the two expressions we get a {\it necessary} condition for the vacuum state $\ket{g,c}$ to be a quasiclassical state
\begin{align}
\frac{\delta W_{loc}[g,c]}{\sqrt{|g|}\delta g^{\mu\nu}(x)}\approx T^{hydro}_{\mu\nu}(u^\lambda(x))\label{preEE}
\end{align}

\subsection{The most general form of $W_{loc}[g,c]$}

The local functional $W_{loc}[g,c]$ gives the fundamental expression for the expectation value of the stress tensor on the left hand side of \ref{preEE}. This is a local approximation to the full nonlocal expression. In this section we use this fact to find the most general form of $W_{loc}$.

The full nonlocal expression for the expectation value of the stress tensor cannot be obtained as a variational derivative of some nonlocal functional like the Schwinger functional because the dependence of the expectation value on the sources is causal while, f.e, taking derivatives of $W_{Schwinger}[g,c]$ yields Feynman, and not the causal propagators. Furthermore, as was mentioned previously, functional derivatives  of the Schwinger functional correspond to matrix elements of the stress tensor between different states (the in- and out- vacua), and not to an expectation value. However, one can utilize the Schwinger functional to compute the expectation values if one follows the Keldysh-Schwinger prescription of computing expectation values of local operators.

The arguments of the rest of this subsections are valid if one considers directly the expectation values, but for the sake of maximal simplicity of the explanation we will pretend as if $W_{loc}[g,c]$ were a local approximation to the Schwinger functional $W[g,c]$.

Thus the question we intend to answer is: under what conditions on the theory and on the sources the nonlocal Schwinger functional $W[g,c]$ is approximated by a local expression $W_{loc}[g,c]$.

There is an obvious obstacle to the existence of such an approximation -- the conformal anomaly. The conformal anomaly is the fundamental property of any QFT. At the level of the Schwinger functional $W[g,c]$ it is the statement that under Weyl transformation of the metric $g_{\mu\nu}\to g^\Omega_{\mu\nu}\equiv e^{-2\Omega}g_{\mu\nu}$ with a parameter $\Omega(x)$ there is a corresponding local transformation of the other sources $c\to c^\Omega$ such that the Schwinger functional is almost invariant, but only up to a local functional ${\cal A}$:
\begin{align}
W[g^\Omega,c^\Omega]=W[g,c]+{\cal A}[g,c,\Omega].
\end{align}

The expression for the conformal anomaly ${\cal A}[g,c,\Omega]$ changes with a change of the Schwinger functional differing by a local functional $W'=W+W_{loc}$\footnote{In generic backgrounds there is no reason to prefer one over another.}, but it can never be shifted away to zero due to the presence of homologically nontrivial terms in ${\cal A}$ -- terms that cannot be obtained with a Weyl transformation of a local functional. The well known example of such a term in ${\cal A}$ is the square of the Weyl tensor
\begin{align}
a\int d^4x\sqrt{|g|}W_{\alpha\beta\gamma\delta}W^{\alpha\beta\gamma\delta}\subset{\cal A}[g,c,\Omega].
\end{align}
Here $a$ is a constant called the a-anomaly coefficient. There are many more homologically nontrivial terms especially when one one switches on the sources $c^a(x)$.\footnote{See \cite{Osb, Ratal} and references therein.}

Thus in the space $D(g,c)$ of the sources the local approximation can only exist in the corners where the homologically nontrivial terms in the conformal anomaly ${\cal A}$ are negligible. One can always assume that this is achieved by having huge coefficients in front of terms in the local expression $W_{loc}[g,c]$ which are the same as homologically nontrivial ones. For example, one could have $10^{50}\int d^4x\sqrt{|g|}Weyl^2\subset W_{loc}[g]$ so that the anomaly can be safely ignored. This is not the case we are interested in. We want to find a natural situation without this huge hierarchy.  This is our main guiding principle for finding the most general expression for $W_{loc}$.

From the existence of the a-anomaly one can conclude immediately that in $W_{loc}$ there cannot be a term which is the square of the Weyl tensor, at least if it does not have a very large coefficient.

We argued earlier that for the existence of the quasiclassical approximation the sources $c^a(x)$ have to be very close to constants. In other words, there must exist a conformal frame in which they are such. Such a frame is fixed up to constant Weyl transformations with $\Omega(x)=\Omega=const$. The classical/effective expression for the stress tensor does not break the scaling symmetry, thus in the following we work in this scaling class of frames. This scaling symmetry is nevertheless approximate as some terms in the conformal anomaly survive the constant scaling -- like the $a-$anomaly itself, so we should not scale 'too far away'.

The most general form of $W_{loc}$ consistent with the remaining scaling symmetry, dimensional analysis and diffeomorphism invariance is
\begin{eqnarray}
W_{loc}[g,c] = \Lambda(c)\int d^4x\sqrt{|g|}+ M^2(c)\int d^4x\sqrt{|g|}R+\alpha(c)\int d^4x \sqrt{|g|}R^2\nonumber\\
+\beta(c)\int d^4x\sqrt{|g|}R_{\mu\nu}R^{\mu\nu}+\gamma(c)\int d^4x\sqrt{|g|}W_{\alpha\beta\gamma\delta}W^{\alpha\beta\gamma\delta}.\label{Wgen}
\end{eqnarray}

Here $\Lambda, M$ are function of the coupling constants $c^a$ of mass dimension four and one, respectively, and $\alpha,\beta,\gamma$ are dimensionless functions of $c^a$. As we have noted, $\gamma$ is allowed only if it is extremely large.

The formula \ref{Wgen} is the most general one which has a regular limit of all coupling constants set to zero (so the UV CFT limit). In particular, the existence of such a limit forbids negative powers of $M$ from entering the expression. This limit must exists because nothing singular happens to the stress tensor in the effective/classical approximation.

We stress that the local approximation is not a local expansion of a nonlocal expression -- that would be singular in the UV CFT limit. Neither does it correspond to integrating out any massive modes which would render the resulting expression singular in the zero limit of their masses.

Next, the Schwinger functional does in fact depend nontrivially on the renormalization scale $\mu$: $W=W[g,c;\mu]$. Thus in principle the local approximation $W_{loc}$ could depend on it as well, so that arbitrary powers of the curvature would enter in the expression \ref{Wgen} multiplied by the inverse powers of $\mu$, f.e.:
\begin{align}
\frac{1}{\mu^{10}}\int d^4x\sqrt{|g|}R^6\subset W_{loc}[g,c;\mu].
\end{align}
However, such terms break the remaining scaling symmetry under which all sources transform but the renormalization scale $\mu$ is left intact. Thus they do not appear on the right hand side of \ref{Wgen}.

Now we substitute this most general expression into the equation (\ref{preEE}) and use the correspondence limit -- the limit of the flat metric. In this background the vacuum $T_{\mu\nu}(particles, fields)$ is trivial. For the case of hydrodynamics: $u^\mu(x)\equiv0, T^{hydro}_{\mu\nu}(x)\equiv0$. This immediately sets the cosmological constant $\Lambda$ to zero!

This conclusion deserves to be emphasized: in the quasiclassical regime the coupling functions become nearly constant on the typical scales of variation of the metric. In the limit when they are exactly constant the cosmological constant is also exactly zero, so the function $\Lambda(c^a(x))$ depends only on spacetime derivatives of the coupling functions $c^a(x)$. Because they are only approximately constant (although to a very high precision), the cosmological constant is not exactly zero, but has a tiny nonzero value! Here we should add that we neglect a small nonlocal correction to the expression of the stress tensor as a function of the metric. However, in the Minkowski case, it is zero as well, so it cannot mimic a nonzero Cosmological constant.

Let us now return to imposing constraints on $W_{loc}[g,c]$ stemming from homologically nontrivial terms in the conformal anomaly. With coupling functions $c^a(x)$  switched on there are many of them \cite{Osb,Ratal}. In particular, there will be terms of the form $\int d^4x\sqrt{|g|}M^2R$ for the relevant deformation $M\widehat{\cal O}_M$ thus making the presence of this term in $W_{loc}[g,c]$ inconsistent without a huge coefficient. So the interpretation of $M$ as the parameter of a relevant deformation is problematic.

If we are to keep this term in $W_{loc}$ we have to find its interpretation which forbids its appearance in the expression for the anomaly. This is not difficult to do: the conformal anomaly appears as the coefficient of the identity operator $\widehat 1$ in the expression for the trace of the stress tensor (\ref{trace}). This means that the anomaly is an operator property which does not depend on a choice of states. Correspondingly, a vacuum expectation value $v$ of a local operator $\widehat{\cal O}(x)$ cannot enter directly into the expression for the anomaly! But it can enter the expression for $W[g,c]$ because it does depend on states. The vev $v$ is typically a limit of two coupling constants when they are both sent to zero:
\begin{align}
v=\lim_{c_{1,2}\to0}\frac{c_1}{c_2}.
\end{align}

This gives another way to see that the vev cannot enter the anomaly by itself since the anomaly is regular in the limit $c_2\to0$. It can only enter, f.e., in the form $c_1=c_2v$ if $c_1,c_2$ are not taken to the extreme zero limit.

Thus, the term $\int d^4x\sqrt{|g|}M^2R$ is absent in the anomaly if $M$ (or $M^2$) is a vev and not the value of a relevant coupling! The corresponding couplings $c_1$ and $c_2$ do not have to be sent to zero -- that would be unnatural -- they just need to be very small, so there is a naturally large hierarchy between the term in $W_{loc}$ and the homologically nontrivial term in the conformal anomaly.

Now we argue that the coefficients $\beta(c)$ and $\gamma(c)$ in the expression (\ref{Wgen}) for $W_{loc}$ are zero in the limit of zero coupling functions (but a finite vev) by considering the correspondence limit. Consider the case of the flat metric and constant sources $c$.

Let us take the UV CFT limit by sending all coupling functions to zero but keeping the vev $M$ and a nontrivial metric $g_{\mu\nu}(x)$. We get
\begin{align}
W_{loc}[g,c] =\int d^4x\sqrt{|g|}M^2R.\label{WCFT}
\end{align}

Because the vev $M(x)$ is not a coupling functions, its spacetime variation is not forbidden by the validity of the quasiclassical approximation --  to the diffeomorphism-invariance equation \ref{diffinv} it contributes only the innocuous trace of the stress tensor $\frac{\partial M}{M}\widehat T^\mu_\mu(x)$ which is a 'good' operator. Now, in the flat metric with constant $M$ the effective matter tensor is identically zero. So,
\begin{align}
\frac{\delta W_{loc}}{\sqrt{|g|}\delta g^{\mu\nu}(x)}|_{g=\eta}=0.
\end{align}
But this must be true in all other conformal frames  -- since now, with all coupling constants set to zero, this will not introduce their spacetime dependence destroying quasiclassicality of $\widehat T_{\mu\nu}$.  The only way to achieve this is to modify (\ref{WCFT}) to make it Weyl-invariant. There is a unique such modification:
\begin{align}
W_{loc}[g,M] =6\int d^4x\sqrt{|g|}\{g^{\mu\nu}(x)\partial_\mu M(x)\partial_\nu M(x)+\frac{1}{6}M^2(x)R(x)\}.\label{Wlocm}
\end{align}

At the same time, there is no conformal completion of $\beta$ and $\gamma$ terms, thus they are zero in the limit $c^a=0$.

This limit is thus useful to reproduce the dependence of $W_{loc}$ on $M(x)$. Nevertheless this situation is extremely fined tuned because it requires taking $c_1,c_2$ to zero and doing this very precisely. What we expect in the real world is small but nonzero values of the pair $(c_1,c_2)$.

Let us relate their space-time variation to the variation of $M(x)$:
\begin{align}
\frac{\delta M}{M}=\frac{\delta c_1}{c_1}-\frac{\delta c_2}{c_2}.
\end{align}

Assume for concreteness that the mass dimension of $c_1$ is two and that of $c_2$ is one.

Then
\begin{align}
\delta M(x)\propto \frac{M}{c_1}\delta c_1(x)=\frac{1}{c_2}\delta c_1(x).
\end{align}

Given smallness of $c_2$ we immediately conclude that $\delta M\gg\delta c_1$ -- the absolute variation of the 'Planck' functions is observable on scales much smaller than that of the fundamental constants $c_1$. From this we immediately get the leading contribution to the cosmological 'constant':
\begin{align}
\Lambda(x)\propto\partial M_P\partial M_P.\label{CCM}
\end{align}

In words, the leading contribution to the cosmological 'constant' comes from spacetime variation of the Planck 'constant'! Both are tiny in quasiclassical region of the Universe.

Let us review the logic here. There is an hierarchy in spacetime variation of all the coupling functions $c^a$ and the vev $M(x)$: we have to keep the coupling functions very close to constant values in order to preserve the validity of the quasiclassical approximation. At the same time there is no such constraint on the vev $M(x)$ -- it only contributes to the operator equation for the (non)conservation of stress tensor (\ref{diffinv}) a harmless trace of the stress tensor. Thus, although it seems that using the conformal invariance of (\ref{Wlocm}) we could set spacetime variation of $M(x)$ to zero, we do not have this freedom -- that would make coupling constants into coupling functions destroying quasiclassicality -- more precisely, the choice of the stress tensor corresponding to that frame is bad -- the tensor operator is not quasiclassical.

The spacetime variation of $M(x)$ is what it is. The full equation taking into account this spacetime variation is

\begin{align}
\frac{\delta}{\sqrt{|g|}\delta g^{\alpha\beta}(y)}6\int d^4x\sqrt{|g|}\{g^{\mu\nu}(x)\partial_\mu M(x)\partial_\nu M(x)+\frac{1}{6}M^2(x)R(x)\}=T^{pts,flds}_{\alpha\beta}(y).\label{EEM}
\end{align}

This is a necessary condition for the Universe to be quasiclassical, but of course it is not sufficient.

From the spatial homogeneity of the Universe we get that the gradient of the Planck 'constant' is timelike. Let us then align the time direction with this gradient. Then
\begin{align}
\Lambda\propto g^{00}(t)\left(\frac{d M}{dt}\right)^2=\left(\frac{\Delta M}{\Delta\eta}\right)^2,
\end{align}
where we introduced conformal time $\eta$. Because the time variation of the Planck 'constant' is very slow, we can approximate it with a constant slope
\begin{align}
\frac{\Delta M}{\Delta\eta}\approx const.
\end{align}

The quantity that we denoted as $\Lambda$ is actually the 'vacuum energy density' or, more precisely, the dark energy and not a cosmological 'constant' or a cosmological function. They actually have different equations of state: they both produce negative pressure, but the signs of their contribution to energy are opposite. Approximately
\begin{align}
T^\Lambda_{\mu\nu}=diag \Lambda(1,-1,-1,-1),\qquad T^{\dot M_P}_{\mu\nu}=\left(\dot M_P\right)^2(-1,-1,-1,-1).
\end{align}
That the variation of $M_P$ produces negative pressure means that it acts as 'antigravity' and so is a form of Dark Energy, but whether this equation of state is even remotely realistic (after taking into account time dependence of energy and the Planck mass), we do not know at the moment.

Let us look at the amount of negative pressure produced.  It can be approximately related to the standard definition of the effective cosmological constant $\Lambda_s$ through the relation
\begin{align}
\Lambda = M^2(t)\Lambda_s(t).
\end{align}
We see that the standard effective cosmological constant becomes time-dependent. Let us find its value at a moment $\eta$ of conformal time:
\begin{align}
\Lambda_s(\eta)=\frac{\Lambda}{M^2(\eta)}=\left(\frac{\Delta M}{M(\eta)\Delta\eta}\right)^2=\left(1-\frac{M(0)}{M(\eta)}\right)^2\left(\frac{1}{\eta}\right)^2.
\end{align}

Assuming that there has been a noticeable but not large change in the Planck constant since the 'beginning' of the Universe so that the first parenthesis is of order unity we get the relation between the cosmological constant and the conformal age of the Universe
\begin{align}
\Lambda_s\propto\frac{1}{\eta_0^2}.
\end{align}

Now, given the change of the cosmological 'constant' in time, the measurement of $\Lambda_s$ which gave the value $\Lambda_s\approx 10^{-122}t_P^{-2}$ did not measure the present value of the 'constant', but its value several billions years ago or some averaged value. Moreover, the dark energy in our modification of Einstein equations is not a Cosmological constant or a Cosmological function -- it is not proportional to the metric. In any case, this should not affect significantly the order of the magnitude estimate.
For the present conformal age of the Universe $\eta\approx 40 byrs\approx10^{61}t_P$ we get the crude estimate for the value of the cosmological 'constant':
\begin{align}
\Lambda_s\propto 10^{-121}t_P^{-2}.
\end{align}

This is not a bad coincidence with the actual value for a crude estimate! What this implies is that there has been a change of order one in the Planck mass $M_P$ on the time scale of the age of the Universe. This does not mean that at present time or even on the timescale of the age of the universe it changed significantly. We do know that the past 10 billion years or so the quasiclassical approximation works very well. However, as we go farther back in time, the Universe becomes more quantum and so we expect the variations of $M_P$ to be larger closer to the big bang or right before it.

Let us estimate the present time $t_0$ variation of the Planck mass:
\begin{align}
a^{-2}(t_0)\left(\frac{dM_P(t_0)}{dt}\right)^2=\Lambda_s(t_0)M^2_P(t_0)\longrightarrow\frac{\dot M_P}{M_P}=a(t_0)\sqrt{\Lambda_s(t_0)}=a(t_0)10^{-61}M_P\nonumber\\
\approx a(t_0)10^{-2}byr^{-1}.
\end{align}

This is close to the upper bound on the time variation of the Newton constant $G_N$ from observations of type $Ia$ supernova \cite{Ia}:
\begin{align}
\frac{\dot M_P}{M_P}\le 10^{-1}byr^{-1}.
\end{align}
Those bound are obtained under certain assumptions on the standard cosmology (C$\Lambda$DM) and do depend on the model. Since the bound is not significantly lower than our estimate, we do not regard it as ruling out our interpretation of the dark energy immediately.

That was a large-scale (intergalactic) bound on the variation of the Planck mass. It turns out that bounds obtained locally, in the solar system, or in the Milky Way galaxy in the current epoch are much lower \cite{Ia}:
\begin{align}
\frac{\dot M_P}{M_P}\le 10^{-4}byr^{-1}
\end{align}

However, the picture of this section is not complete. We have not introduced all the players in the game. What we can conclude at the moment is that on large scales the picture is not obviously ruled out, while at the local scale it is definitely inadequate. Thus, if there is hope to respect the local bounds, the new players have to be local.

While we did consider global quasiconserved currents as quasiclassical variables on the same footing as the stress tensor, so far we ignored them, but in the real world we do have such currents -- for example the baryon symmetry current.

So we continue in the next section with their inclusion in the picture.

Finally, an important comment is in order at this point. The expression (\ref{Wlocm}) looks like the action for the Brans-Dicke theory with a parameter $\omega=4/6$ (the conformal case) while the experiment has ruled out values of $\omega<40000$.

One should be absolutely clear about the conceptual difference between our interpretation of $W_{loc}[g,M]$ and Brans-Dicke theory. In the latter $M(x)$ is a dynamical field with Equations of Motion obtained from the variation of $W_{loc}[g,M]$ with respect to  $M(x)$. In our approach $M(x)$ is not dynamical, it is what it is. The validity of the quasiclassical approximation requires only that it is very close to a constant. The only reason the metric becomes effectively a dynamical variable is because the stress tensor is a quasiclassical variable for which there are two expressions -- the fundamenta/cosmological/microscopic one and the effective/local/macroscopic ansatz, and the two expressions must match. As for $M(x)$, it does not correspond to any quasiclassical variable -- correspondingly there is no classical ansatz to match to (other than identical proportionality to the trace of the stress tensor which does not add any constraints). In conclusion,  $\omega=4/6$ is not ruled out in our context.

We would like to further emphasize that for this reason the matching equation (the generalized Einstein Equations) (\ref{EEM}) determines the metric unambiguously on scales small compared to the scale of variation of $M(x)$ (so small compared to the age of the Universe), but on longer scales exactly due to the fundamental absence of an equation of motion for $M(x)$ one cannot compute the metric unambiguously from the distribution of matter.

Finally, we would like to make an important point. In our approach the dark energy is not a cosmological constant or a cosmological 'function' -- the variation of $W_{loc}[g,M]$ with respect to the metric gives the dark energy contribution which is not just proportional to the metric -- there is the term
\begin{align}
\partial_\mu M\partial_\nu M\subset\frac{W_{loc}[g,M]}{\sqrt{|g|}\delta g^{\mu\nu}}.
\end{align}

 Thus we get an explanation of the coincidence $t^2_U\propto\frac{1}{\Lambda_s}$, but also the explanation of the sign of the effective Cosmological constant (as far as the pressure is concerned)

\section{A sketch of inclusion of other (quasi)conserved currents.}

This section has a preliminary character and more of an announcement of ideas, but we think it is worth announcing them for completeness of the picture.

So far we have assumed that the stress tensor is the only quasiconserved 'current' (more precisely it gives quasiconserved currents after contraction with quasi-Killing vectors) in the theory. In this section we investigate what happens when we include additional such currents $\widehat J_{\mu}(x)$. We will confine the discussion to a single current, but the generalization to several is trivial.

The motivation for including additional quasi-conserved currents is of course the fact that in the real Universe there are several: the baryon symmetry current $B$, the lepton symmetry current $L$ and the electric current $j$. It maybe that the first two give rise to a $B-L$ current in a GUT theory which is conserved to a higher degree. Here we restrict our attention to the current whose conservation is broken the least. Let us call this current the baryon current since in the real world it will probably contain $B$.

The motivation to singling out such a current is the following. If the current conservation is strongly broken, then it ceases to be a good quasiclassical variable by the same argument that we used for the stress tensor in the presence of spacetime varying coupling functions: the conservation constraint $\nabla_\mu\widehat J^{\mu}=c\widehat{\cal O}(x)$ is contaminated by the 'bad' local operator $\widehat{\cal O}(x)$ who is not quasiclassical in the sense that its approximate eigenstate at an initial moment $t_0$ quickly ceases to be such at later times. Through the non-conservation equation it transfers this property to the current $\widehat J$ itself. Correspondingly, to prevent further contamination of the stress tensor one has to put the spacetime variation of the source $A_{\mu}(x)$ to zero just like with other coupling functions.

If, on the other hand, the currents is conserved exactly, there is no reason to set the variation of $A_{\mu}(x)$ to zero -- it can have variations comparable to the variations of the metric without any problems for the validity of the quasiclassical approximation.

Obviously, then, if the current is broken only slightly we get a hierarchy of scales: $L(g_{\mu\nu})\ll L(A_\mu)\ll L(c)$ -- the scale of spacetime variation of the metric is much smaller than the scale of variation for the source $A_{\mu}(x)$ of the quasi-conserved current $J_{\mu}$ which is itself much smaller than that of coupling constants.

Consider the effect of this background source on probe charges. The diffeomorphism invariance condition (\ref{diffinv}) gives the equation
\begin{align}
\nabla^\mu \widehat T_{\mu\nu}(x)-F_{\mu\nu}\widehat J^\mu+A_\nu\nabla^\mu\widehat J_{\mu}=0,\label{Adiff}
\end{align}
where $F_{\mu\nu}\equiv\partial_\mu A_\nu-\partial_\nu A_\mu$. The matrix element of this operator equation between the in- and out-vacua in asymptotically Minkowski case can be obtained from the diffeomorphism invariance of the Schwinger functional $W[g,A;\mu]$, with the action of the diffeomorphism algebra $\delta x^\mu=\xi^\mu(x)$ on the 1-form $A=A_\mu dx^\mu$ being
\begin{align}
\delta_\xi A={\cal L}_\xi A= i_\xi dA+d(i_\xi A).
\end{align}

As with the stress tensor, the quasiconserved current being a quasiclassical variable, in addition to the microscopic expression for its expectation value $J_\mu(x)=J_\mu[g(z),A(z);x]$, one gets a macroscopic one
\begin{align}
J_\mu(x)\approx J_\mu[u^\lambda(z),x]
\end{align}
which should be matched. In the local approximation this matching condition becomes
\begin{align}
\frac{\delta W_{loc}[g,A]}{\sqrt{|g|}\delta A^\mu(x)}\approx J_{\mu}^{hydro}(u(x)).
\end{align}

As is well known \cite{Wald}, for the case of $A\equiv0$ the covariant conservation of the stress tensor for dust particles is equivalent to the geodesic equations of motion due to the continuity relation $\nabla^\mu u_\mu=0$
\begin{align}
0=\nabla^\mu (mu_\mu u_\nu)=m(\nabla^\mu u_\mu)+mu_\mu\nabla^\mu u_\nu=mu_\mu\nabla^\mu u_\nu
\end{align}

Then in the presence of a nonzero vector field/form $A$, neglecting the non-conservation of $\widehat J_\mu$, from the equation (\ref{Adiff}) and with the hydro ansatz $J_\mu=eu_\mu$ one gets the equation
\begin{align}
mu_\mu\nabla^\mu u_\nu = F_{\mu\nu}u^\mu,\label{EM}
\end{align}
which is nothing but the usual equation of motion of a relativistic test charge with mass $m$ and charge $e$ in the gravitational and 'electromagnetic' fields.

We must find the dependence of $W_{loc}[g,A]$ on $A$. In this paper we will not do this. In particular, because it requires a careful analysis of conformal anomalies in the case of a slightly non-conserved current, and this is work in progress. But we do consider a particular case here. The case of an exactly conserved one was considered, in particular, in \cite{Ratal}. In that case the anomaly contains the usual Maxwell kinetic term $F^2$ for $A_\mu$ -- correspondingly it cannot appear in $W_{loc}[g,A]$ since the local approximation only works in regimes when such terms are neglected.

This means that if the baryon number is conserved exactly, no local approximation $W_{loc}[g,A]$ exists for $A_\mu\ne 0$ unless there is a huge hierarchy appearing as a huge coefficient in front of the Maxwell term in $W_{loc}[g,A]$. If there is no such hierarchy, one would have to solve the full nonlocal Einstein equations. The downside to this possibility, in addition to complexity of the task, is that this apparently breaks quasiclassicality which is unlikely, so maybe there is a hierarchy with nonzero $A_\mu$.

In any case, at this time we do not have even any equations for the background source $A_\mu(x)$. Nevertheless, let us see what modifications to the motion of galactic matter they would produce. Namely, let us estimate, under what circumstances the Dark Matter could potentially be explained by a nonzero sources $A_\mu(x)$.

To reproduce the galaxy rotation curves at large distances from the galactic bulges, which we at this stage we assume are caused by the field $F$ generated by the bulge we need to have a 'magnetic' field directed along the rotation axis zero 'electric' field. We will have a Taylor expansion for the magnetic field
\begin{align}
B(\rho)=B_0+\frac{B_1}{\rho}+\frac{B_2}{\rho^2}+... .
\end{align}

In the standard Electromagnetism the first two terms are absent. The first one would correspond to a linear rise in velocity of the test charge with the distance in circular motion around the bulge, and the second term -- to the constant linear velocity according to the nonrelativistic approximation to the 'Lorenz' equation (\ref{EM})
\begin{align}
m\frac{v^2}{\rho}=evB(\rho),\qquad\qquad v(\rho)=\frac{e}{m}(B_0\rho+B_1).
\end{align}

An important point is that the 'charge' being the baryon charge, gives the same ratio $\frac{e}{m}$ for all test particles. If it actually contains a mix of lepton charge $e=b-l$, due to the small mass of leptons compared to baryons, the variation in the ratio is negligible. Thus the motion of all test particles will be the same due to the absence of antimatter -- a point we will return to later.

It is trivial to find that the form $A=\alpha d\phi$ where $\alpha=const$ is the required solution with fields
\begin{align}
B=\frac{\alpha}{\rho}d\phi,\qquad\qquad F_{0i}=0.
\end{align}

Thus we do get the constant linear velocity of rotation
\begin{align}
v(\rho) = \alpha =const.
\end{align}

As our final point, let us return to the case of the nonzero 'electric' field. In its presence the baryons and anti-baryons would move in opposite directions if their own produced field is not strong enough to prevent this. Since the dynamics is unlikely to be described by the Maxwell theory, this is a possibility. If this baryonic 'electric' field were present in the early Universe, it could lead to the separation of matter and anti-matter on (meta-)cosmological scales.

\section{Resume}

We sum up our proposal. We assume that the wave function of the Universe $\ket{\Psi}_U$ is the simplest one one can have in QFT -- the only parameters that enter are the parameters of a QFT: the metric $g_{\mu\nu}(x)$ and a set of coupling functions $c^a(x)$. More precisely, it is the equivalence class $(g_{\mu\nu},c^a)/Weyl$ of pairs $(g_{\mu\nu}, c^a)$ related by Weyl transformations so that only the causal structure is physical. In this situation the stress tensor $\widehat T_{\mu\nu}(x)$ has the expectation value
\begin{align}
{}_U\bra{\Psi}\widehat T_{\mu\nu}(x)\ket{\Psi}_U=\bra{g,c}\widehat T_{\mu\nu}(x)\ket{g,c}=T_{\mu\nu}[g,c](x)
\end{align}
is a nonlocal functional of the metric and the coupling functions, which depends on them causally. There are many such stress tensor which transform one to another under changes of conformal frames. The important point is that they mix with other local operators and with the identity operator. The latter is a manifestation of the conformal/Weyl anomaly.

The above is the fundamental/global description of the expectation value, but there is also an effective/local one -- in terms of particles and fields:
\begin{align}
T_{\mu\nu}[g,c](x)=T_{\mu\nu}(particles, fields)(x).
\end{align}

This relation may naively be viewed as a nonlocal version of Einstein equations of General Relativity, because they seem to related the metric and the matter. This is not so. For this interpretation to be valid one has to have the effective density matrix $\rho_{eff}(particles, fields)(x)=tr_{environment}\ket{g,c}\bra{g,c}$ computed from the wave function of the Universe by taking the partial trace of the environment to be strongly peaked around a (averaged over macroscopic volumes) $\widehat T_{\mu\nu}$-eigenstate. Only in this case can one interpret the above relation as a nonlocal version of Einstein Equations.

The persistence in time of such a peak around a $\widehat T_{\mu\nu}$- eigenstate corresponds to the quasiclassical situation. We argued that a necessary condition for the latter is quasi-constancy of the coupling functions. More precisely, there must exist a conformal frame in which coupling functions $c^(x)$ are nearly constant in spacetime.

One may consider, for example, a hydrodynamical situation where $T_{\mu\nu}(particles, fields)(x)=T_{\mu\nu}(hydro)(u(x))$. Due to the constancy of sources $c^a$ the condition of diffeomorphism invariance is reduced to the covariant conservation of the stress tensor. Then one gets the usual hydrodynamical equation
\begin{align}
\nabla^\mu T_{\mu\nu}(u(x))=0.
\end{align}

In the case of dust $T_{\mu\nu}(u(x))=\rho(x)u_\mu(x)u_\nu(x)$ one gets the motion along geodesics for the metric $g_{\mu\nu}(x)$. With fixed initial metric and $u^\mu(t_0)$ one finds the four-velocity $u^\mu(x)$ as a functional of the metric $g_{\mu\nu}(t>t_0)$, and compares to the expression on the left hand side which depends only on the metric. Naturally, one does not get a match -- with fixed initial values this will only happen for a special metric $g_{\mu\nu}(t>t_0)$. In this way this constraint relates the geometry and matter.

On may expect that for a quasiclassical Universe the nonlocal functional $T_{\mu\nu}[g(y),c](x)$ can be approximated by an expression local in the metric. There is an obstacle of having this approximation naturally -- without huge coefficients -- the conformal anomaly. We argued that this obstacle is overcome if the UV CFT contains a vacuum in which the conformal symmetry is broken. That is, a couple of coupling functions $c^1(x)$ and $c^2(x)$ are weakly turned on so that the resulting vev $v(x)=\frac{c^1(x)}{c^2(x)}$. Sending these coupling functions to zero while keeping the ratio fixed would be extremely fine-tuned so they are small but nonzero. Furthermore, each of them is forbidden from having any noticeable variations in spacetime due to the quasiclassicality condition, but their ratio, the vev $v(x)$ can vary much more.

We argued that in this situation the local approximation to the LHS of (\ref{}) is given by the variation of the expression
\begin{align}
W_{loc}[g,c]=\int d^4x\sqrt{|g|}(v^2R+6g^{\mu\nu}\partial_\mu v\partial_\nu v)
\end{align}
with respect to the metric. The cosmological constant is exactly zero in the absence of spacetime variations of coupling functions because in the limit of Minkowski spacetime there are neither particles no fields in the standard vacuum, so $\Lambda \eta_{\mu\nu}=T_{\mu\nu}(particles, fields)=0$!  At the same time, there are no equations of motion for the vev $v$, so it is not a version of Brans-Dicke theory! The vev $v(x)$ is a free parameter. The only restriction on it is that its spacetime variations are tiny. Moreover, there is the condition $v^2R\gg Weyl^2$ which is the condition of negligibility of the conformal anomaly. At this point we identify the vev  with the Planck mass $v=M_P$.

Due to the spatial homogeneity of the universe the gradient $\partial_\mu v$ has to be timelike -- which automatically gives the positive contribution to the Dark Energy.

The effective cosmological constant
\begin{align}
\Lambda_s(t)=6a^{-2}(t)\left(\frac{d M_P}{dt}\right)^2\propto 10 \dot M_P^2
\end{align}
where we took into account that the cosmological constant is measured at redshifts around one, and $\dot M_P$ is the average rate of change of Planck mass over the last several billion years corresponding to redshifts $z\propto 1-2$. From the cosmological observations of type Ia supernovae one has the constraint $\dot M_P\le10^{-61}$ in Planck units which gives rise to the bound
\begin{align}
\Lambda_s(z\propto 1-2)\le 10^{-121}
\end{align}
in agreement with the measured value $\Lambda\propto 10^{-122}$. This interpretation of the Dark energy is also appealing because it explains the approximate relation between the cosmological constant $\Lambda$ and the age of the Universe $t_U$:
\begin{align}
\Lambda\propto\frac{1}{t_U^2}.
\end{align}

However, from local measurements of $\dot M$ in the solar system or the galaxy one gets a tighter bound on the present rate of change of $M_P$: $|\dot M_P|\le 10^{-63}$ leading to $\Lambda_s\le 10^{-125}$ which is three orders of magnitude less than the measured value.

But perhaps, one should not consider the local bound on the time variation of $G_N$ as ruling out our expression for the dark energy for two reasons. First of all, the time variation could be smaller in the recent time than on cosmological time scales at which the Dark Energy was observed\footnote{One may wonder if a decelerated decay of $M(t)$ (forward in time) and thus of $\Lambda_s(t)\propto a^2(t)\left(\dot M\right)^2$ in the past could mimic the inflation.}. Secondly, this diminishing in the time variation could be an effect local in space -- there could be something like screening off the time derivative $\dot G_N$ by local effects on galactic scales, especially considering that the gravitational dynamics on this scale is not well understood.

In conclusion, we feel that one should only consider the bounds obtained in the same setup in which the Dark Energy itself was measured -- on cosmological time scales. This bound $|\dot M_P|\le 10^{-61}$ in Planck Units gives rives to the bound on the Dark Energy density $\rho\le 10^{-122}$ marginally consistent with the measured value $\Lambda\propto 10^{-122}$. Moreover, in \cite{GB} a polynomial of dependence of Newton's constant $G_N$ on the redshift $z\le2$ was fitted to the experimental data
\begin{align}
G_N(z)=G_N^0(1-0.01z+0.3z^2-0.17z^3)
\end{align}
giving rise to the estimate $\dot M\propto 10^{-61}$ in Planck units during last few Gigayears.

Note that so far we assumed that the local effect from the time variation of $M_P$ is bigger than the nonlocal correction to the Einstein equations which is small, but not zero since fundamentally $T_{\mu\nu}[g,c](x)$ is a nonlocal functional of the metric. We had no reason to assume this hierarchy -- it could well be the other way around, so that the Dark Energy is dominated by the contribution of the tiny nonlocal part of $T_{\mu\nu}[g,c]$. It is interesting to note that some work with mimicking the effect of the Cosmological Constant by a nonlocal modification of Einstein Equations has been done recently (\cite{MM,AB} and references therein). We are not experts in this field to estimate the success of this approach, but given that these nonlocalities are fundamental consequences of our view of Einstein Equations, their effects on cosmology is a primary goal of our future work.

We conclude with an important speculative remark suggested by our approach. The scale at which the effective density matrix $\rho_{eff}(matter, fields)$ is strongly peaked around a (smeared) $\widehat T_{\mu\nu}(x)$ eigenstate -- the scale at which matter and geometry start to be related by Einstein equations -- is not the scale at which effective determinism appears! This is due to the fundamental fact that in QFT the metric/causal structure is a free parameter. In quasiclassical situations it is related to the distribution of matter/energy, so if between an initial and final time at this scale the situation remains 'clean' from 'contamination' of the microscale (the scale at which the reduced density matrix is not strongly peaked around a $\widehat T_{\mu\nu}$ eigenstate) and simple enough at all intermediate times so that Einstein equations can be solved, then one gets the final metric and matter distribution unambiguously from the initial data. If there is such a 'contamination', although both at the initial and the final times the geometry and matter are related by Einstein equations, the best one can hope for is a probabilistic prediction for the metric. Since the metric is a classical 'field', in a sense the probabilities are classical, and the full quantum state is never reduced/updated and always evolves unitarily\footnote{For arguments in favor or compatibility of unitary evolution with 'collapse', see \cite{Omnes}.} in the Schrodinger picture according to the Schrodinger equation with a time-dependent Hamiltonian. The time dependence comes from the time-dependence of the metric $g_{\mu\nu}(t)$. The probabilities appear from the indeterminacy in the metric/causal structure.

We emphasize that unitary time evolution $\widehat U$ does not imply the existence of a conserved Hamiltonian $\widehat H$. In the presence of spacetime dependent sources (including the metric $g(t)$), the unitary evolution is given by
\begin{align}
\widehat U(t_2,t_1)=T\exp{\left(i\int_{t_1}^{t_2}\widehat H(g(t))dt\right)}.\label{Ut}
\end{align}

\section{More on Quantum Mechanics}

Our analysis in the preceding sections prompts us to entertain the following speculation about the structure of Quantum Mechanics: there is an additional Quantum Mechanical property of our Universe, in addition to the standard ones, so that roughly their list is:
\begin{enumerate}
\item Unitary time evolution of the state vector.
\item Causality, locality.
\item (???): additional property.
\end{enumerate}
We did not list the wave function reduction axiom nor any mentioning of measurements because 1) measurements are just particular examples of quantum mechanical evolution (or they should be), 2) the additional property is supposed to eliminate the wave function reduction axiom and allow the Unitary evolution to be universal. We conjecture that this additional property together with the other two leads to the fundamental fact: macroscopic physical subsystems of the Universe almost never exist in superpositions of (averaged over macroscopic volumes) $\widehat T_{\mu\nu}$- (and thus position-) eigenstates. Let us state it more precisely: in the limit $N\to\infty$, the unitary production of subsystems of $N$ degrees of freedom in superpositions of $\widehat T_{\mu\nu}-$ eigenstates is suppressed.

There is no room to accommodate this additional property in elementary Quantum Mechanics, but the real world is not described by an elementary quantum mechanics -- it is described by a quantum mechanical theory which is at least as complicated as Quantum Field Theory. We assume that it is actually a QFT, but with certain restrictions stemming from this property.

The obvious objection to its existence is exactly the situation of a measurement of a microstate in a superposed state
\begin{align}
\frac{1}{\sqrt{|a|^2+|b|^2}}\left(a\ket{+}+b\ket{-}\right),
\end{align}
such that
\begin{align}
U\ket{0}\ket{+}=\ket{P}\ket{+},\nonumber\\
U\ket{0}\ket{-}=\ket{M}\ket{-},\label{U12}
\end{align}
where $U$ is the operator of the unitary evolution of the isolated system 'measuring device+microsystem', and the macrostates of the measuring device $\ket{P}$ and $\ket{M}$ are macroscopically distinct -- by the position of a macroscopic pointer.

We would like to consider the following situation in this experimental setup
\begin{align}
U\ket{0}\frac{1}{\sqrt{|a|^2+|b|^2}}\left(a\ket{+}+b\ket{-}\right)=\ket{P}\ket{+}.
\end{align}
The first objection is that $U$ cannot be unitary because the information about the relative phase of the initial microstate is lost. This is easy to refute: the real measuring device is a macrosystem with many degrees of freedom, not just the pointer, where this information can be preserved.

The second objection is that the unitary operator $U$ is linear, thus the result of its action on the superposed state must be the sum of two terms in (\ref{U12}), so let us return to (\ref{U12}) to consider it more carefully.

The two initial states are parts of the Universe with the wavefunction $\ket{g_{\mu\nu}(x)}$ located in different spacetime regions. According to our conjecture, the macrostates $\ket{M}\ket{-}$ and $\ket{P}\ket{+}$ are approximate $\widehat T_{\mu\nu}$-eigenstates, and thus, being different position-eigenstates, they create different geometries around them. Thus, starting with the same initial geometry $g_0$ corresponding to the initial macrostate $\ket{0}$, the final geometries being different, the unitary operators in the two situations being functionals of the metric are in fact different, so we really should write
\begin{align}
U(g_0,g_P)\ket{0}\ket{+}=\ket{P}\ket{+},\nonumber\\
U(g_0,g_M)\ket{0}\ket{-}=\ket{M}\ket{-},\label{U12c}.
\end{align}
It looks like the Unitary operators of evolution know about the microstate: if it is $\ket{+}$, then $U(g_0,g_P)$ acts, if it is $\ket{-}$, then $U(g_0,g_M)$. Thus is not what happens: the two operators have the same functional dependence on the metric since they correspond to identical measuring devices. It is the metric between the initial and final states that 'knows' about the initial micro-state, but in our Universe, the latter is obtained from $\ket{g_{\mu\nu}(x)}$ by taking the partial trace, and thus these local microstates are functionals of the metric in the causal past of their location.

So the metric during the measurement 'knows' about the metric in its causal past. There is nothing wrong with this. In fact, this is essentially what the statement of applicability of a local approximation of $T_{\mu\nu}[g,c](t,{\bf x})$ boils down to: in this approximation the causal expectation value is 'factorized' through a local ansatz
\begin{align}
T_{\mu\nu}[g,c](t,{\bf x})\approx M_P^2G_{\mu\nu}(t,{\bf x})
\end{align}
in terms of the metric $g_{\mu\nu}$ and its derivatives at point $(t,{\bf x})$, so that the causal dependence of the expectation value of the stress tensor on the metric in $J_-(t,{\bf x})$ is realized through a causal dependence of the metric $g_{\mu\nu}(t,{\bf x})$ at point $(t,{\bf x})$ on the metric in $J_-(t,{\bf x})$.

Now, it is obvious what will happen in the case of the superposition of the two microstates -- the total system will evolve with still another unitary operator which, to avoid the highly suppressed macroscopic superposition will be either $U(g_0,g'_P)$ or $U(g_0,g'_M)$ generically where metrics $g'_P$ and $g'_M$ are macroscopically the same as $g_P$, $g_M$, respectively, but differ from them micro-/meso-scopically.

We should note at this point that the approximation of a unitary evolution of the system 'measuring apparatus+microsystem' may not by itself be consistent. This approximation works for isolated systems, but exactly because the measuring apparatus is macroscopic, it might have non-negligible, and maybe even essential for the consistency of our picture, gravitational interactions with the outside world. But in our picture (nonlocal) Einstein equations are exactly the consistency condition, so proving inconsistency of our scheme with unitary picture of the measurement is potentially inconsistent by itself. The same argument applies to the situation of spreading of an approximate position-eigenstate of a macroscopic object.

The conclusion is that to be consistent with the additional QM property of the Universe, the spacetime on which QFT lives is not arbitrary: the value of the metric $g_{\mu\nu}(x)$ at a spacetime point $x$ cannot be arbitrary with respect to $g_{\mu\nu}(J_-(x))$, but it is only constrained by the past $g_{\mu\nu}(J_-(x))$, and not determined by it completely. For example, in the case of the superposed microstate there are two consistent macro-options for the future metric.

To sum up, we see that this property has to constrain the spacetime manifold on which the QFT lives (probably leaving intact the rest of QFT structure), but it is not uniquely defined -- it can 'fork' in the sense that at certain moments the metric has the freedom to go one direction or another. This is where probabilities enter the theory at the fundamental level. The evolution is always unitary with respect to some spacetime metric, but the metric itself is not uniquely defined by the initial data.

The question that we try to answer next is: why does the consistent assignment of probabilities of future metrics have to follow the Born rule?

We suggested above that the indeterminacy in the  quantum mechanical Universe comes from unpredictability of future geometries, which can be predicted only probabilistically if at intermediate times there is interference from the microscopic world like during the measurement of microscopic systems in the state $(\ket{+}+2\ket{-})/\sqrt{5}$ whose two outcomes correspond to the readings $\ket{P}$ and $\ket{M}$ of a macroscopic measuring device, and thus to two geometries $g(P)$ and $g(M)$. The question is why the consistent assignment of probabilities $p(g(P))$ and $p(g(M))$ should correspond to the Born rule, that is $p(g(P))=1/5$, $p(g(M))=4/5$.

First, one has to keep in mind that to talk about probabilities one has to keep records of the outcome of each experiment which is a physical macroscopic process changing geometry. For example, at the end of each experiment one puts a ball of mass $m=1kg$ in a basket if the result is $\ket{M}$, and for the other outcome one does not do anything. Then there is the corresponding Energy operator for the basket $\widehat T_{00}=m\widehat N_M$, where $\widehat N_M$ is the number operator for the balls in the basket.

The rest, at an heuristic level, is an old idea of Hartle and probably someone else: for a very large number $N$, the state
\begin{align}
\ket{\Psi}_N=\frac{1}{\left(|a|^2+|b|^2\right)^{N/2}}(a\ket{M}+b\ket{P})^N\label{InSt}
\end{align}
is an approximate eigenstate of the number operator $\widehat N_M$ -- the number operator for $\ket{M}$, with eigenvalue $N|\alpha|^2$.

In our context, then, having 5000 copies of the experiment, the final state of the basket is the approximate eigenstate of the energy operator $\widehat T_{00}$ with the eigenvalue $1000kg$:
\begin{align}
\widehat T^{final}_{00}\ket{1000kg}=1000kg\ket{1000kg}.
\end{align}

Thus the corresponding final geometry around the basket is known with probability close to one to be $g(1000kg)$. Correspondingly, the consistent assignment of probabilities for the geometries $g(M)$ and $g(P)$ corresponds to the Born rule: $p(g(M))=1/5$, $p(g(P))=4/5$.

Let us clarify the above argument. We suggested that for macroscopic bodies for which gravity works, the reduction is in fact a unitary process described by the time-dependent Schrodinger equation with the evolution operator $U(g(x))$.

Now if we consider a very large number $N$ of measuring devices each of which is coupled initially to identical microstates $(\ket{+}+2\ket{-})/\sqrt{5}$, then the evolution of the collection of $N$ systems will be with the unitary operator
\begin{align}
U_{tot}=U(g_1(x))\otimes U(g_2(x))\otimes ... \otimes U(g_N(x)),\label{Utot}
\end{align}
where the individual geometries $g_i(x)$ of measuring devices $MD_i$ are not known, so can be treated probabilistically. Then one can expect that for very large $N$ the total operator (\ref{Utot}) will exhibit some universality for reasons similar to the usual Central limit theorem. Indeed, each evolution operator is a time-dependent unitary operator $U(g_i(t))$, but if we neglect the difference between the $T-$exponent and the simple exponent, then the total Hamiltonian for $U$ will be universal in the large $N$ limit: $\widehat H_{tot}=N\widehat H$, where $\widehat H$ does not depend on the set of geometries $\{g_i\}$. We expect the same to be valid in the time-dependent case
\begin{align}
U_{tot}\to U(universal)
\end{align}
in the limit $N\to\infty$ {\it as long as we do not ask fine-grained questions and only are interested in coarse-grained features}, like in the CLT if one only measures the sample average. Then, the coarse-grained evolution being unitary as well, we would have a quasi-conserved coarse-grained operator $\widehat N$ which corresponds to the sample average times $N$ in the CLT. Its initial value obtained by acting on the initial state (\ref{InSt}) which is its approximate eigenstate and contains superpositions gives $N|a|^2$. Because this number is approximately conserved, this is also the final value at the end of the experiment which corresponds to $1000$ balls in the basket and thus to the assignment of probabilities according to Born.

Now we want to conjecture that the additional quantum mechanical property is already contained in our proposal. Namely, it is the form of the Universal wave function $\ket{g,c}$ which only depends on the metric and coupling constants. So, this property boils down to the 'duality' relation
\begin{align}
\bra{g,c}\widehat T_{\mu\nu}(x)\ket{g,c} = T_{\mu\nu}(pts(x),flds(x))
\end{align}
which so far we understood as just an equality of the fundamental/global description of the expectation value of $\widehat T_{\mu\nu}(x)$ and classical effective/local description in terms of particles and fields.

However, there is no reason to view this 'duality' in such a narrow sense. Indeed, the natural step is that we can understand it as the equality
\begin{align}
\bra{g,c}\widehat T_{\mu\nu}(x)\ket{g,c} = tr(\widehat T^{p.,f.}_{\mu\nu}(pts(x),flds(x))\rho_{pts, flds})\label{Qmatch}
\end{align}
where everything on the right hand side is in terms of quantum particles and (quantum) fields -- the trace is over their Hilbert space, and $\rho_{eff}$ is in terms of their positions, momenta and so on. Furthermore, not only the expectation values are supposed to coincide but also some number of correlators of $\widehat T_{\mu\nu}$ in the two representations.

 So we track the evolution of large-scale variables -- the coarse-grained or collective variables like positions of pointers, positions of centers of masses of macro-bodies, averaged densities, etc.

To better explain the next step, let us return to the classical situation for a moment, and consider the Einstein equations for a collection of dust particles, for concreteness:
\begin{align}
M^2_PG_{\mu\nu}(x)=mu_\mu(x) u_\nu(x).\label{hmatch}
\end{align}
From our point of view, this is a consistency condition for the pair $(g_{\mu\nu}(x), u_\lambda(x))$. How does one find a consistent pair starting with an initial pair ${g_0,u_0}$? One starts with a trial metric $g_{\mu\nu}(t,{\bf x})$ and solves the stress-tensor covariant conservation equation with the dust ansatz:
\begin{align}
\nabla^\mu(mu_\mu(x) u_\nu(x))=0,
\end{align}
thus obtaining the pair $(g_{\mu\nu}(t,{\bf x}),u_\lambda[g(y)])$ as a functional of the trial metric which one substitutes back into the ansatz $mu_\mu(x) u_\nu(x)$ to get the functional of the metric on the right hand side of (\ref{hmatch}). Obviously, this will not be equal to the left hand side for any trial metric. In the absence of any other factors, only a single pair should survive the matching.

Now, the same logic as with the hydro approximation applies: starting with an initial metric $g_0$ and initial $\rho_{pts,fds}$ we calculate the right hand side on the time interval $(t_0,t_f)$ for a trial metric $g(t)$ which coincides with $g_0$ at $t_0$, so that the right hand side becomes a functional of $g(t)$, but there is no reason for it to coincide with the left hand side! Indeed, we expect that only a measure zero subset of all trial metrics survives the matching. So now we formulate our conjecture:

{\it For a sufficiently complex initial metric $g_0$ (and its past values), the set of coarse-grained metrics that survive the matching (\ref{Qmatch}) corresponds to the course-grained density matrices whose diagonal entries evolve towards all zeros except one almost unit diagonal entry, while off-diagonal elements remain small. The resulting unitary evolution reproduces the final state obtained from the usual collapse axiom. The corresponding surviving coarse-grained geometries (as is clear from the discussion above) have the structure of a branching tree.}

We summarize our suggestion:

\begin{enumerate}
\item Any path from the top to the bottom of the tree, that is, any chosen evolution of geometry in time, thus exhibits the 'appearance' of non-unitary collapses of wave-function for the coarse-grained variables, but in fact for any such metric the evolution of the entire Universe is unitary\footnote{with a spacetime-dependent metric, so there is not a conserved Hamiltonian (\ref{Ut})}, and probabilities come from unpredictability of the future metric at any point on the 'tree'. The time arrow in the tree was introduced when we required the expectation value of $\widehat T_{\mu\nu}(x)$ to be a causal functional of the metric so that it depended only on the metric in the causal past of $x$ (including $x$ itself).
\item We should emphasize, that the evolution is unitary for any single path, and only one of them is realized in Nature (the red one in the picture) -- there is no branching of the wave-function, no Everett's parallel worlds! Of course, the metric always changes smoothly, which is not reflected in the tree diagram below.
\item Moreover, the evolution being always unitary for some metric, the appearance of 'collapse' is an emergent phenomenon: it emerges with increased complexity of the metric!
\item Finally, it emerges on macro scales (so for coarse-grained variables) together with General Relativity. On smaller scales one has a nonlocal version of it in the form (\ref{Qmatch}).
\end{enumerate}

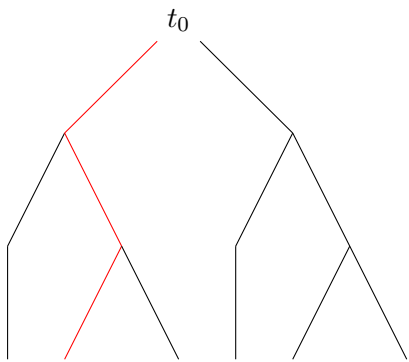
\begin{tikzpicture}[level distance=1.5cm,
  level 1/.style={sibling distance=3cm},
  level 2/.style={sibling distance=1.5cm}]
  \node {{$t_0$}}
    child {[red] {}
      child {[black]{}
      child {}
      }
      child {[red]{}
      child{}
      child{[black]}}
        }
    child {{}
    child {child{}{}}
    child {child{}child{}{}
    }
    };
\end{tikzpicture}

We conclude with a few remarks on the apparent collapse of the local wavefunction/density matrix during measurements to clarify our conjecture.

Consider an experimental setup in which a particle is created at a spacetime point $x_0$, after which it is free to take two paths to end up in a position-superposed state at a later moment:
\begin{align}
U(T)\ket{x_0}=\frac{1}{\sqrt{|\alpha|^2+|\beta|^2}}\left(\alpha\ket{x}+\beta\ket{y}\right).
\end{align}
There is a macroscopic particle detector $M_x$ at position $x$. According to our suggestion, depending on the geometry the result of the measurement will be either the macroscopic state $\ket{0}_M$ of the device corresponding to 'no particle' at $x$, or $\ket{1}_M$ corresponding to 'the particle is at $x$' -- in each case obtained unitarily but with different metrics. The probabilities for the two metrics are, respectively, $|\beta|^2/\sqrt{|\alpha|^2+|\beta|^2}$ and $|\alpha|^2/\sqrt{|\alpha|^2+|\beta|^2}$.

The statement of the usual collapse axiom (and our conjecture) is stronger than this: not only will the final state of the device be in either of the two states, but the state of the particle itself will be correlated with the state of the measuring device, so that the two outcomes are
\begin{align}
\ket{0}_M\ket{y},\nonumber\\
\ket{1}_M\ket{x}.\label{fc}
\end{align}
Whatever the dependence of the interaction between the measuring apparatus and the particle on the metric is, it is local, and hence, it cannot reproduce the two final alternatives. Indeed, the result, if it were due only to the interaction between the two would be the following:
\begin{align}
\ket{0}_M\ket{y},\nonumber\\
\ket{1}_M\left(\ket{x}+\gamma\ket{y}\right),\label{fnc}
\end{align}
where $\gamma\ne0$, and $\ket{y}$ is left since it does not interact with the measuring device located at a spacelike-separated point $x$.

However, we do not say that the collapse (for all practical purposes) comes from the local interaction between the measuring device and the particle -- it comes from the consistency condition (\ref{Qmatch}) on the metric, and both $\ket{y}$ and the apparatus $M_x$ are coupled to the metric, so there is no surprise that their final states (after 'collapse') are correlated exactly because the future metrics at positions $x$ and $y$ are correlated since they share the common past at $x_0$.

More precisely, the metric around $x_0$ does differ slightly depending on the outcome of the measurement. There is nothing spooky about this -- the different metrics correspond to the solution of the matching equation with different future boundary condition -- one, corresponding to the boundary condition of particle detection (the future state of the apparatus is $\ket{1}$), the other -- to the alternative outcome ($\ket{0}$). We expect the difference to be very small prior to some time-like surface which contains the measuring event, which very abruptly grows on this surface. The amount of abruptness is controlled by the ration of detector's mass to the Planck mass and maybe some other parameter.

We should stress that although very tiny, there is some nonzero amount of 'collapse' prior to the contact with the measuring device.

One might object that the difference in the final alternatives (\ref{fc}) and (\ref{fnc}) is only in the microstates, which has little influence on the metric, but one should keep in mind that measuring apparatuses are essentially amplifiers of the microworld, so that, the evolution being unitary, the information about the relative phases of the initial microstate is preserved in the internal degrees of freedom of the measuring device in a mesoscopic way. In the second option (\ref{fnc}), the information of the phase is still in the microstate, and thus, due to the non-cloning theorem, it is not in the internal degrees of freedom of the device. This makes the two options different mesoscopically. In particular, the metrics in the two cases differ mesoscopically. Hence, it makes a significant difference for the satisfaction of the consistency condition (\ref{Qmatch}), and we may expect that the first situation (\ref{fc}) passes the match, while the second (\ref{fnc}) does not.

\end{document}